\documentstyle[11pt,paspconf]{article}

\begin{document}

\title{Terminal Velocity Infall in QSO Absorption Line Halos}

\author{Robert A. Benjamin\altaffilmark{1}}
\affil{Dept of Physics, University of Wisconsin-Madison}

\altaffiltext{1}{Previously at Minnesota Supercomputer Institute, 1200 Washington Ave., Minneapolis, MN 55415}

\def\cm2{{\rm cm^{-2}}}
\def\kpc{{\rm kpc}}
\def\kms{{\rm km~s^{-1}}}
\def\cm3{{\rm cm^{-3}}}

\begin{abstract} 

We explore the hypothesis that clouds detected in quasar absorption
line systems are falling at a terminal velocity toward the center of
high redshift gaseous galactic halos.  Since both the ionization level
and terminal velocity of halo clouds increase with increasing distance
from the central galaxy, velocity resolved profiles of highly ionized
gas are predicted to have a greater width than low ionization gas. A
line of sight passing through the center of gaseous halo (an idealized
damped Ly alpha system), yields low ionization absorption at the
velocity of the galaxy, flanked by high ionization on either
side. Reasonable halo parameters yield total velocity extents for C IV
of $\Delta v_{C IV}=100-200 ~{\rm km~ s^{-1}}$, in agreement with
several observed systems.  The remaining systems may
better described by the rotating disk model of Prochaska \& Wolfe
(1998).  Finally, observational tests are suggested for verifying or 
falsifying the terminal velocity hypothesis for these systems.

\end{abstract}


\section{Do Quasar Absorption Line Clouds Fall at Terminal Velocity?}

The spectra of distant quasars typically exhibit numerous absorption
lines which arise in discrete intervening gas clouds. Many of these
clouds, particularly those producing metal absorption lines, are
thought to be embedded in the gaseous disks and halos of ``normal''
high redshift galaxies (Bahcall \& Spitzer 1969).  In a previous
paper, it has been argued that drag may dominate the motions of neutral
clouds observed in the gaseous ``halo'' of our Galaxy, and that these
clouds fall at or near terminal velocity toward the Galactic plane
(Benjamin \& Danly 1997, BD97). Generalizing these results to a
spherical system using reasonable profiles for enclosed mass and gas
density, the terminal velocity is

\begin{equation}
v_{T}(r)=25.3~{\rm km~s^{-1}}~ C_{D}^{-1/2}~v_{r,100}~n_{f,-2}^{-1/2}~N_{19}^{1/2}~r_{10}^{(n-1)/2}~,
\end{equation}

where $C_{D} \approx 1$ is the drag coefficient, $N_{19}$ is cloud
total hydrogen column density integrated along the direction of cloud motion
in units of $10^{19}~ \rm{cm^{-2}}$. The mass enclosed by radius
$r_{10}=r/10~\kpc$ in units of $10^{10}~M_{\sun}$ is $M_{10}(r)=2.3
v_{r,100}^{2}r_{10}$, where $v_{r,100}=v_{r}/(100~ \kms )$ is the
rotational velocity associated with this mass distribution. The gas
density exterior to some fiducial radius, $r_{f}$, is
$n_{h}(r)=n_{f,-2}(r/r_{f})^{-n}$, where $n_{f,-2}$ is the halo
density at $r_{f}=10~\kpc$ in units of $10^{-2}~\cm3$.  Here we use
n=2 (c.f., Mo \& Miralda-Escude 1996). We implicitly assume the
absorption arises due to a collection of discrete clouds intervening
along a line of sight passing through a more tenuous gaseous
halo. This is consistent with the tendency for absorption lines to
break into multiple components when observed with higher
resolution. For simplicity the halo is assumed to be spherical and
steady-state and the clouds falling radially inward.  For a line of
sight passing through the halo with impact parameter, $r_{imp}$, the
velocity spread of absorption is $\Delta
v=2v_{T}(R_{i})[1-(r_{imp}/R_{i})^{2}]^{1/2}$, where $R_{i}$ is the
maximum radius at which absorption due to an ion $i$ occurs.  Allowing
for a reasonable range of input parameters, i.e. $0.5 <v_{r,100} <2.5$
(c.f., Casertano \& van Gorkom 1991), $0.01<n_{f,-2}< 10$ (c.f., Mo \&
Miralda-Escude 1996), $0.1 < N_{19} <10$ (c.f., Dickey \& Lockman
1990), and $3 h^{-1}<R_{i,10}<10h^{-1}$ (Lanzetta 1993), equation (2)
satisfies the observed velocity spread of $100 <\Delta v <200~ \kms $ (c.f., Lanzetta \& Bowen 1992; Churchill, Steidel, \& Vogt
1996; Lu, Sargent, \& Barlow 1996).

\section{First Test: Is There a Correlation Between Ionization and Velocity?}

Given the latitude in the above parameters, one can say that observed
velocities spreads are consistent with, but not conclusive evidence
for, terminal velocity infall.  Consideration of the velocity profile
as a function of ion may provide additional support for the terminal
velocity model in some systems. Statistics of quasar absorption line
systems indicate that at a given redshift, the absorption cross
section increases with level of ionization of absorber ion.  At $z \cong 1.5$, $R_{Mg~II} \cong 66 h^{-1}~\kpc$, while
$R_{C~IV}=98 h^{-1}~\kpc$. Although the cross
section depends on the geometry of absorber and
cosmological model (c.f.  Lanzetta 1993, Steidel 1993), the general
result that $R_{C~IV}>R_{Mg~II}>R_{DLA}$ (DLA=''Damped Lyman Alpha'')
is robust.

The simplest (but not only) explanation of this correlation is that it
indicates an increase in the ionization parameter of the absorbing
clouds with radius. Bergeron \& Stasinska (1986) have shown that the
ionization parameter, $\Gamma=\phi/cn_{c}$, inferred from comparison
of photoionization models to observational data is consistent with
constraints on the extragalactic ionizing radiation field, $\phi \cong
(0.9 \pm 0.5) \times 10^{6}~ {\rm H~ionizing~photons~s^{-1}~cm^{-2}}$
for $z=2-4$ (Cooke, Espey, \& Carswell 1996), and reasonable cloud
densities, $n_{c}$. But regardless of the origin of this correlation,
if the terminal velocity model is appropriate, there should be a
second correlation between ionization level and velocity.  Highly
ionized clouds in the diffuse outer halo fall faster, while more
neutral clouds in the inner halo fall more slowly.

Quantifying these expectations requires adapting a model for the
ionization mechanism and structure of the clouds. Using version 84.12a
of CLOUDY (Ferland 1993), we model the clouds
as constant density slab photo-ionized by the intergalactic radiation
field of Madau (1992) with total hydrogen column density
$\log~N_{H}=19$ and metallicity $\log (Z/Z_{\sun})=-2$, assuming 
a uniform radiation field everywhere in the halo. In order to
link the cloud ionization, which depends on $n_{c}$, to the
cloud dynamics, which depends on $n_{h}$, we assume initially that
clouds maintain a constant ratio $\chi=n_{c}/n_{h}$.  We calculate the
absorption column making no correction for the inclination of the
absorbing cloud relative to the line of sight.  Such a correction
could be important but depends on the detailed geometry, morphology,
and internal density structure of falling clouds which is poorly
understood (BD97).
 
In Figure 1, we illustrate a few examples of absorption depth for C IV
1549 \AA~ and Fe II 1608 \AA~ as a function of velocity. We assume a
``base'' case with the following parameters: $\phi_{6}/\chi_{2}=1.0$,
$n_{f,-2}=1$, $n=2$, $v_{r,100}=2.0$, $r_{imp}=0$, $N_{19}=1$, and
$\log (Z/Z_{\sun})=-2$. The parameter varied is noted in each
panel. In each case, the spatial integration is restricted to
$R_{max}=300 \kpc$, and the absorption envelope is artificially
truncated at that point. The photoionization models were carried out
with a resolution of only 0.25 dex in ionization parameter, the
resulting profiles of Fe II have an unphysical cusp at low velocities.
Since we have not specified a model for the location of individual
absorbers, nor how the density of absorbers depend upon radius,
these are not {\it bona fide} synthetic absorption line profiles, but
rather define an absorption ``envelope'' into which more detailed
models will fit.

As expected, the absorption envelopes show a
neutral low-velocity core, and two high ionization wings. If the halo
density structure flattens interior to some radius (as suggested by Mo
\& Miralda-Escude 1996) the velocity width of the neutral core will
decrease. The envelopes here are comparable to the 
data taken for selected systems: $z_{abs}=2.8268$ in Q 1425+6039 
(Lu et al 1996) and $z_{abs}=2.8443$ in HS 1946+7658 (Tripp et al ) are two examples.

\begin{figure}[!htb]
\plotfiddle{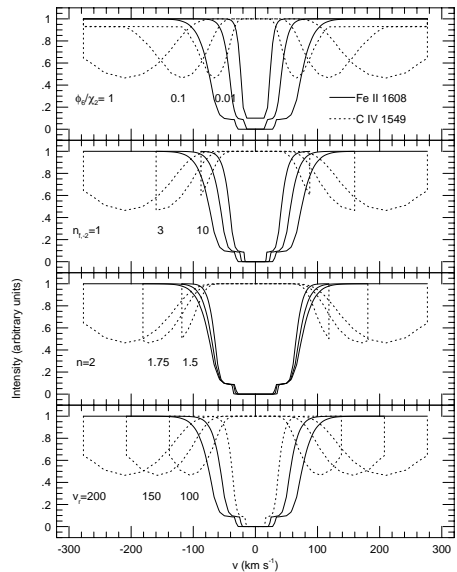}{75truemm}{0}{45}{45}{-150}{-80}
\caption{Calculated ``absorption envelopes'' for a line of sight with impact
parameter $r_{imp}=0$. The fiducial parameters are listed in the text; the four panels 
show the effect of varying the halo-to-cloud density conversion
factor, $\phi_{6}/\chi_{2}$ [top panel], the overall density level,
$n_{f}$ [second panel], the slope of the density profile, $n$,[third panel], and the overall galaxy 
mass, characterized by $v_{r}$ [bottom panel]. Comparison of these
profiles to observations could constrain
the physical parameters of QSO absorption line halos.}
\end{figure}

\section{Second Test: Is There a Correlation Between Component Column Density 
and Velocity Spread?}

The three principal classes of motions expected to occur in gaseous
halos which have been compared with various subsets of data are radial
infall (Lanzetta \& Bowen 1992), rotation(Prochaska \& Wolfe 1997),
and velocity dispersion (Charlton \& Churchill 1996 ; McDonald \&
Miralda-Escude 1998).  {\it The principal difference between our
proposed dynamical model and these other models is that drag-dominated
motions depend upon the column density of the cloud}, whereas the
other classes of motion have been assumed to be independent of cloud
parameters.  We expect that the more an absorption component is offset
in velocity from the assumed centroid velocity, the higher the column
density of the cloud \underline{or} the greater the distance from the
kinematic center \underline{or} both. From equation (1), the velocity
spead of components should go as $\Delta v \propto \sqrt{N}$, assuming
that the gaseous halo is uniformly filled with clouds of varying
column density. It should be emphasized that the relevant parameter is
the \underline{total} column density of the feature; this can only be
obtained from the data by applying a ionization model to convert
column density of individual ions into the total column. Quantitative
predictions are currently being prepared.

\section{Deciding on a model}

All models of quasar absorption line systems must necessarily make a
large number of assumptions in order to make predictions that can be
directly be compared with observations. Many of these assumptions are
not directly related to the kinematics of individual components, but
will affect the prediction for observations. Variations in
metallicity, radiation density, filling factor, or mean absorber
column density are all assumed here to be constant with radius, since
we lack firm theoretical prejudices or observational guidance to tell
us exactly how such quantities should vary. But such variations
probably do occur, and will affect our ability to observationally
discriminate between the primary classes of motion.

Another difficulty is that depending upon the orientation and history
of the absorption system, rotation, infall, and velocity dispersion
may all be occurring simultaneous, and matching the profile requires
tuning the parameters of all three types of motions simultaneously.
Nearly edge-on disks will have profiles dominated by rotation while
face-on disks or spheres will have profiles more dominated by vertical
or radial infall, and the structure of the profiles as a function of
ion will depend upon the impact parameters.

Given the large number of parameters that are necessary to compare
models of QSO absorption line systems to the data, final adoption of
any model will ultimately be motivated by a combination of
satisfactory agreement with data and physical plausibility. Building a
model of high redshift halos requires constructing the interstellar
medium of galaxies from first principles, so one test of the physical
plausibility of a model is whether it is agrees with behavior in our
own Galaxy. Rotation and velocity dispersion have been characterized
in our own Galaxy for a long time. Drag-dominated infall has
only recently been proposed for our Galaxy (BD97), and the status of this
proposal is summarized in Benjamin (1999).  Since it has only been 
mentioned in passing for QSO absorption line halos (Mo 1995), it seems
worthwhile to consider the potential importance of this type of
motion. After all, at the time galaxies were first formed, gaseous
infall, terminal velocity or otherwise, must have been the predominant
type of motion.

\acknowledgments

I would like to thank Todd Tripp, Blair Savage, Don Cox, Limin Lu, and
Tom Jones for useful discussions, Nathan Ferry for his hospitality
during my visit, and Gary Ferland for the use of the
photoionization code CLOUDY. This work was supported by NASA grants
NAGW-3189 and grant NAG5-3155, and the Minnesota Supercomputer
Institute.



\begin{references}

Benjamin, R.A. 1999 in The High Velocity Cloud Workshop, eds. B.K. Gibson \& 
M.E. Putman, ASP Conference proceedings, in press

Benjamin, R.A., \& Danly, L. 1997, ApJ, 481, 764

Bergeron, J., \& Stasinska, G. 1986, A\&A, 169, 1

Bahcall, J.N., \& Spitzer, L., Jr. 1969, ApJ, 156, L63

Casertano, S., \& van Gorkom, J.H. 1991 AJ, 101,1231

Charlton, J.C., \& Churchill, C.W. 1996 ApJ 465, 631

Churchill, C.W., Steidel, C.C., \& Vogt, S.S. 1996 ApJ, 471, 164

Cooke, A.J. ,  Espey,B., \&  Carswell, B. 1997 MNRAS, 284, 552

Dickey, J.M., \& Lockman, F.J. 1990, ARA\&A, 28, 215

Ferland, G. J. 1993, University of Kentucky Department of Physics and
Astronomy Internal Report

Lanzetta, K.M. 1993, in The Environment and Evolution of Galaxies, eds. Shull, J.M., \& Thronson, H.A., Jr. (Kluwer: Dordrecht) 237

Lanzetta, K.M. \& Bowen, D.V. 1992 ApJ, 391, 48

Lu, L. et al 1996 ApJS, 107, 475

Madau, P. 1992 ApJL, 389, 1

McDonald, P. \& Miralda-Escude, J. 1998 ApJ, submitted

Mo, H.J. 1995 in QSO Absorption Lines, ed. G. Meylan (Springer) 445

Mo, H.J., \& Miralda-Escude, J. 1996 ApJ, 469, 589

Prochaska, J.X. \& Wolfe, A.M. 1997, ApJ, 487, 73

Steidel, C.C. 1993, in The Environment and Evolution of Galaxies, eds. Shull, J.M., \& Thronson, H.A., Jr. (Kluwer: Dordrecht) 263

Tripp, T.M., Lu, L., \& Savage, B.D. ApJS 102, 239


\end{references}
\end{document}